\documentclass[
 reprint, superscriptaddress,
nofootinbib,
 amsmath,amssymb,
 aps,
]{revtex4-2}
\usepackage{graphicx}
\graphicspath{ {/paper/}}
\usepackage{dcolumn}
\usepackage{bm}
\usepackage{slashed}
\usepackage{amsmath}
\usepackage{enumitem}
\usepackage{blindtext} 
\usepackage[linktocpage]{hyperref}
\usepackage[svgnames]{xcolor}
\usepackage{color}
\hypersetup{
     colorlinks = true,
     linkcolor = magenta,
     anchorcolor = blue,
     citecolor = blue,
     filecolor = blue,
     urlcolor = blue
     }

\usepackage{bbold}
\usepackage[hang,small,bf]{caption}
\usepackage{braket}
\usepackage{tikz}
\usetikzlibrary{backgrounds,fit,decorations.pathreplacing}
\usetikzlibrary{quantikz}
\usepackage{kantlipsum}
\usetikzlibrary{shadows,arrows,calc,backgrounds,fit,shapes,snakes,shapes.multipart,decorations.pathreplacing,shapes.misc,patterns,positioning}
\usepackage{pgfplots}
\usepackage{subcaption}
\begin{document}
\preprint{APS/123-QED}

\title{Quantum Homogenization as a Quantum Steady State Protocol on NISQ Hardware}

\author{Alexander Yosifov}
\email{yosifov.alexander@huawei.com}
\affiliation{Theory Lab, Central Research Institute, 2012 Labs, Huawei Technology Co. Ltd., Hong Kong Science Park, Hong Kong SAR}

\author{Aditya Iyer}
\email{aditya.iyer@physics.ox.ac.uk}
\affiliation{Clarendon Laboratory, University of Oxford, Parks Road, Oxford, OX1 3PU, UK}

\author{Daniel Ebler}
\email{ebler.daniel1@huawei.com}
\affiliation{Theory Lab, Central Research Institute, 2012 Labs, Huawei Technology Co. Ltd., Hong Kong Science Park, Hong Kong SAR}
\affiliation{Department of Computer Science, The University of Hong Kong, Pokfulam Road, Hong Kong SAR}

\author{Vlatko Vedral}
\email{vlatko.vedral@physics.ox.ac.uk}
\affiliation{Clarendon Laboratory, University of Oxford, Parks Road, Oxford, OX1 3PU, UK}

\date{\today}

\begin{abstract}
Quantum homogenization is a reservoir-based quantum state approximation protocol, which has been successfully implemented in state transformation on quantum hardware. In this work we move beyond that and propose the homogenization as a novel platform for quantum state stabilization and information protection. Using the Heisenberg exchange interactions formalism, we extend the standard quantum homogenization protocol to the dynamically-equivalent \texttt{($\mathtt{SWAP}$)}$^\alpha$ formulation. We then demonstrate its applicability on available noisy intermediate-scale quantum (NISQ) processors by presenting a shallow quantum circuit implementation consisting of a sequence of \texttt{CNOT} and single-qubit gates. In light of this, we employ the Beny-Oreshkov generalization of the Knill-Laflamme (KL) conditions for near-optimal recovery channels to show that our proposed \texttt{($\mathtt{SWAP}$)}$^\alpha$ quantum homogenization protocol yields a completely positive, trace preserving (CPTP) map under which the code subspace is correctable. Therefore, the protocol protects quantum information contained in a subsystem of the reservoir Hilbert space under CPTP dynamics.
\end{abstract}

\maketitle



\section{Introduction}
Physical system with the capacity to consistently entangle its degrees of freedom and maintain a coherent quantum state arbitrarily close to a desired target state for time longer than the lifespan of single modes is a prerequisite for the development of noisy quantum devices, with implications to many quantum protocols, \textit{e.g.} long-distance quantum communication \cite{3a}, quantum networks \cite{3b}, quantum cryptography \cite{3}, and quantum sensing \cite{3c}.

So far, many stabilization schemes \cite{feedback1, feedback2, qfeedback, qfeedback2} are inspired by classical control theory and rely on direct feedback loops for carrying quantum feedback procedures, where a \textit{controller} compares the system output with a predetermined target value, and makes adjustments in real time based on the measurement-acquired information. Alternatively, some models utilize open quantum systems \cite{lindblad}, governed by a Lindblad master equation with varying dissipative and Hamiltonian terms, while others \cite{lindblad2} construct stabilizer codes and eliminate the dissipative part altogether.

These approaches, however, are fundamentally hindered by the fact that measurements induce spurious backreactions on the system. Although in this respect weak measurements were considered \cite{weak}, they only provide limited amount of information, while requiring precise control over individual degrees of freedom, thus leading to considerable experimental complexity and hardware overhead \cite{overhead}.

To bypass those shortcomings, reservoir-based dynamical systems have been recently proposed as a viable alternative for a variety of tasks, such as coherent quantum control and state preparation \cite{dissipative,dissipative2,dissipative3}. Here, the repetitive engineered coupling of an input system to a reservoir has already demonstrated utility for the robust entanglement of superconducting qubits \cite{states, qubits}, showing improved state stabilization and prolonged entangled state lifetimes. The appeal of this idea stems from employing the system-reservoir dynamics as a resource for engineering steady states which are available ``on demand" and immune to some errors, thus reducing cost and improving scalability. Although very promising, the main challenge for this method lies in engineering a physical system with dynamics that naturally drives it to a desired steady state without actively controlling individual degrees of freedom.

Motivated by the recent success of such exchange interaction models \cite{8,9,10}, in this work we contribute in this direction by proposing the quantum homogenization protocol \cite{4} as a novel platform for quantum state stabilization. The protocol is a quantum information formalization of state approximation collision models \cite{collision1,collision2} via sequential input-reservoir interactions, modelled by the two-body partial \texttt{$\mathtt{SWAP}$} --- a versatile operation which implements a probabilistic exchange of states between the input system and the reservoir \cite{thermo,thermo1,thermo2, 5, 6, vlatko, vlatko2}. Fundamentally, the repetitive system-reservoir interactions spread the system information across the reservoir, maintaining coherence and entangling its degrees of freedom. Concretely, we substitute the partial \texttt{$\mathtt{SWAP}$} gate with the bipartite \texttt{($\mathtt{SWAP}$)}$^\alpha$ operator \cite{heisenberg}, generated by a Heisenberg exchange interaction, where $\alpha$ is time-dependent coupling parameter, and show that the \texttt{($\mathtt{SWAP}$)}$^\alpha$ is equivalent to the partial \texttt{$\mathtt{SWAP}$}. Upon this insight, we demonstrate the practical applicability of the \texttt{($\mathtt{SWAP}$)}$^\alpha$ quantum homogenization protocol by presenting an experimentally-realizable shallow quantum circuit, consisting of four \texttt{CNOT} gates and six single-qubit gates, which can be natively implemented on current NISQ processors. Then, in the context of the Beny-Oreshkov generalization \cite{oreshkov} of the KL correctability conditions, we argue that the iterative input-reservoir interactions of the proposed \texttt{($\mathtt{SWAP}$)}$^\alpha$ quantum homogenizer encode logical states in a reservoir code subspace via a noisy channel in a correctable form. Hence, a recovery quantum channel exists, such that it \textit{approximately} reverses the encoding process. This demonstrates that the homogenization protocol naturally converges to a desired steady state and protects the encoded information in the reservoir subspace.\footnote{A novel observation was made in \cite{preskill} that quantum reservoirs can protect quantum information against errors of polynomial complexity.}

\section{Quantum homogenization}
Classical reservoir computing is a simple facilitator of neural network models \cite{crc}. It utilizes the dynamics of a fixed reservoir (a collection of artificial neurons or physical degrees of freedom) to process input data without fine-tuning. Recently, that framework was extended to the quantum domain \cite{qrc}, where, through the use of quantum modes, inherently quantum effects (\textit{e.g.} entanglement) can be leveraged.

The quantum homogenizer is a reservoir-based protocol \cite{4, vlatko, vlatko2} with naturally convergent dynamics \cite{convergence}. It takes any input quantum state $\rho_{S}\in\mathcal{H}_{S}$ and applies a series of probabilistic exchange operations with the reservoir degrees of freedom --- $N$ $d$-state quantum systems (\textit{i.e.} qudits, see \cite{qudits, catstates}), each prepared in arbitrary state $\xi^{(i)}=\xi$ $\forall i\in N$, leading to the the density matrix $\mathcal{H}_{R} \ni \xi_{\mathbf{R}}^{\otimes N} :=\xi_{{R}_1}\otimes \xi_{{R}_2}\otimes ... \otimes \xi_{{R}_N}$ after $N$ rounds.\footnote{For numerical simulations demonstrating the relation between state fidelity and the size of $N$, see \cite{anna}.} The exchange dynamics is modelled by the two-body partial \texttt{$\mathtt{SWAP}$} operator $U_{{S}{R}} = \text{cos}\, \eta\, \mathbb{1}_{{S}{R}} + i\, \text{sin}\, \eta\, \texttt{$\mathtt{SWAP}$}_{{S}{R}}$, where $\eta$ is the parameter determining the probability of either leaving the state invariant (by applying the identity operator $\mathbb{1}_{{SR}}$) or exchanging the input state with the given qudit state through the $\texttt{$\mathtt{SWAP}$}$ operation as

\begin{equation}
    \texttt{$\mathtt{SWAP}$}_{{S}{R}} \left( \rho_{{S}} \otimes \xi_{{R}}  \right) \texttt{$\mathtt{SWAP}$}^{\dagger}_{{S}{R}} = \xi_{{S}} \otimes \rho_{{R}} \
\end{equation}
The quantum homogenization protocol then gradually converges the input $\rho_S$ towards $\xi$ as
\begin{equation}
\label{eq:1}
U_{N}^{\dagger}\dots U_{1}^{\dagger}\left(\rho_{{S}}\otimes \xi^{\otimes N}_{\textbf{R}}\right) U_{1}\dots U_{N}\approx\xi^{\otimes N+1}_{{S}\textbf{R}} \ 
\end{equation}
where $U_{k} := U_{{S}{R}_k} \otimes(\otimes_{j\neq k}\mathbb{1}_{j})$ denotes the unitary two-body interaction between the input system and the $k\text{th}$ reservoir qudit.

In the homogenization process, the input state gets gradually mapped from the input space $\mathcal{H}_{{S}}$ onto a subspace of the Hilbert space $\mathcal{H}_{{R}}$ of the reservoir, where due to acting with $U_{{S}{R}}$ (for small $\eta$) the changes to the reservoir have been experimentally shown to be vanishingly small and the state of individual reservoir qudits remains coherent \cite{vlatko, qudits}. Generally, given that at each timestep the input system interacts with a single reservoir qudit, after the first step it evolves as 

\begin{equation}
\label{eq:cpmap}
\rho^{(1)}_{{S}} = \text{tr}_{{{R}_{1}}}\left[U_1(\rho^{(0)}_{{S}}\otimes\xi^{(1)})U_1^{\dagger}\right]
\end{equation}
where its state becomes \cite{4} 

\begin{equation}
\rho_{{S}}^{(1)} = (\text{cos}\,\eta)^{2}\rho_{{S}}^{(0)}+(\text{sin}\,\eta)^{2}\xi+i\,\text{cos}\,\eta\,\text{sin}\,\eta\left[\xi,\rho_{{S}}^{(0)}\right]  
\end{equation}
with the term $[\xi,\rho_{{S}}^{(0)}]=\xi \rho_{{S}}^{(0)} - \rho_{{S}}^{(0)} \xi$ denoting the commutator of the two states, and  $\eta$ being the coupling parameter. Analogously, the state of the first ancilla $\xi^{(1)}\in {R}_1$ reads \cite{4} 
\begin{equation}
\xi^{(1)} = (\text{sin}\,\eta)^{2}\rho_{{S}}^{(0)} + (\text{cos}\,\eta)^{2}\xi + i\,\text{cos}\,\eta\,\text{sin}\,\eta\left[\rho_{{S}}^{(0)},\xi\right] \
\end{equation}
After the input system interacts sequentially with $N$ reservoir qudits, $\rho^{(N)}_{{S}}$ reaches a steady state as\footnote{The steady state properties of the quantum homogenization protocol were extended to continuous variable qudits in \cite{qudits}.}

\begin{equation}
\label{eq:densityoperator}
\rho^{(N)}_{{S}} = \text{tr}_{\bold{R}}\left[U_{N}...U_{1}\left(\rho^{(0)}_{{S}}\otimes\xi^{\otimes N}_{\bold{R}}\right)U_{1}^{\dagger}...U_{N}^{\dagger}\right]
\end{equation}
where $U_{k}$ is given as in (\ref{eq:1}). That is, the dynamics of the quantum homogenizer is captured by an operation $U_{{SR}}$ which asymptotically converges (contracts) all input states towards a desired steady state. In other words, the sequence of dissipative operations asymptotically drives a system in any initial state to the thermal state of the reservoir, which in our case is a superposition of $d$ coherent states \cite{13, raginsky}.\footnote{For a comprehensive treatment of quantum homogenization as a contractive map, see \cite{4}. Note that, compared to qubit models, the engineered coupling of a reservoir with a higher-dimensional quantum system (\textit{e.g.} qudit) has been proven to be an efficient resource for stabilizing the coherent state of the system \cite{highdim}.}

In this respect, as it was proven in \cite{qudits}, the contractivity of $U_{{SR}}$ ensures that $\mathcal{D}(U_{{SR}}(\rho^{(\ell)}_{{S}}\otimes \xi_R)U_{{SR}}^\dag) \leq \gamma_\ell \mathcal{D}(\rho^{(\ell)}_{{S}}\otimes \xi_R)$, where $\gamma_\ell\in [0,1)$, $\forall \ell \geq 1$, and $\mathcal{D}$ is the Hilbert-Schmidt distance \cite{14,15}. Consequently, for $\ell=N$ large enough, $\mathcal{D}(\rho^{(N)}_{{S}},\xi)\leq\delta$, where $\delta \ll 1 $ is a small positive constant. Here, the set of steady states, corresponding to the dynamical process (\ref{eq:densityoperator}), is given by the set of density operators $S(\mathcal{H}_{{R}})$ \cite{states2}. Where for the case of $d$-dimensional qudits, we know from \cite{catstates,steadystate}, that under such engineered system-reservoir coupling the system asymptotically converges to $S(\mathcal{H}_{{R}})$, thus making $\mathcal{H}_{R}$ the asymptotically steady state manifold.

Note that the utilization of such two-body exchange interactions with a dissipative reservoir has already yielded promising results for quantum state stabilization of a parametric oscillator \cite{steadystate2} and quantum data classification \cite{classification}.\\

\section{Heisenberg exchange formalism}
Suppose we substitute the partial \texttt{$\mathtt{SWAP}$} gate $U_{SR}$ with the two-body Heisenberg exchange operator \texttt{($\mathtt{SWAP}$)}$^\alpha$ \cite{heisenberg}, herein denoted as $U^{\alpha}$, where $\alpha$ is the tunable coupling parameter (updated at each timestep) which controls the strength of the interactions between the input and the ancilla \cite{8,9,10}.\footnote{It was shown in \cite{heisenberg} that three \texttt{($\mathtt{SWAP}$)}$^\alpha$ gates are necessary and sufficient for the construction of an optimal quantum circuit. Moreover, the \texttt{($\mathtt{SWAP}$)}$^\alpha$ gates are known to be as efficient as \texttt{$\mathtt{CNOT}$} gates in implementing two-body operations.}

In the following analysis, without loss of generality, we restrict our analysis to the case of qubit messages and reservoir states. The Bell basis for a pair of qubits is defined as

\begin{equation}
\label{eq:Bellbasis}
\begin{aligned}
\ket{\Phi^{\pm}}=\frac{1}{\sqrt{2}}\left(\ket{00}\pm\ket{11}\right)\\
\ket{\Psi^{\pm}}=\frac{1}{\sqrt{2}}\left(\ket{01}\pm\ket{10}\right)
\end{aligned}
\end{equation}
where using (\ref{eq:Bellbasis}), $U^{\alpha}$ can be expressed as \cite{heisenberg}

\begin{equation}
\label{eq:explicit}
\begin{split}
U^{\alpha} = \ket{\Phi^{+}}\bra{\Phi^{+}} + \ket{\Phi^{-}}\bra{\Phi^{-}} + \ket{\Psi^{+}}\bra{\Psi^{+}} \\+ e^{i\pi\alpha}\ket{\Psi^{-}}\bra{\Psi^{-}}  \
\end{split}
\end{equation}
The relation between the partial \texttt{$\mathtt{SWAP}$} gate $U_{{SR}}$ and $U^{\alpha}$ can be seen via the matrix representation

\begin{equation}
\label{eq:matrix}
\begin{aligned}
U_{{SR}} &= 
\begin{pmatrix}
    \cos\eta + i\sin\eta & 0 & 0 & 0 \\
    0 & \cos\eta & i\sin\eta & 0 \\
    0 & i\sin\eta & \cos\eta & 0 \\
    0 & 0 & 0 & \cos\eta + i\sin\eta
\end{pmatrix} \\~\\
&= e^{i\eta} 
\begin{pmatrix}
    1 & 0 & 0 & 0 \\
    0 & \frac{1+e^{\frac{i\pi}{n}}}{2} & \frac{1-e^{\frac{i\pi}{n}}}{2} & 0 \\
    0 & \frac{1-e^{\frac{i\pi}{n}}}{2} & \frac{1+e^{\frac{i\pi}{n}}}{2} & 0 \\
    0 & 0 & 0 & 1
\end{pmatrix}\\~\\
&= e^{i\eta}{U^{\alpha}}
\end{aligned}
\end{equation}
where $-2\eta = \frac{\pi}{n}$, and $\alpha=\frac{1}{n}$. Therefore, it is evident that the $U^{\alpha}$ operator \cite{heisenberg} is equivalent, up to a global phase factor of $e^{-i\eta}$, to the standard partial \texttt{$\mathtt{SWAP}$} $U_{{SR}}$, as defined in \cite{4,vlatko,vlatko2}. Remark that (\ref{eq:matrix}), together with single-qubit rotations, is universal for simulating any unitary two-qubit gate \cite{8}.

\begin{figure}[ht]
\centering
\includegraphics[width=\linewidth]{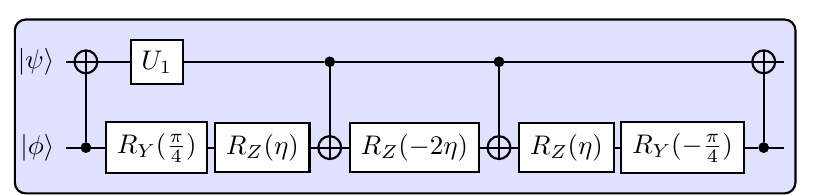}
\caption{Quantum circuit representation of the proposed \texttt{($\mathtt{SWAP}$)}$^\alpha$ quantum homogenizer.}
\label{fig:figure}
\end{figure}

To demonstrate the feasibility of the proposed quantum gate on hardware with native \texttt{CNOT} gates, such as current superconducting NISQ devices, we have decomposed it to a combination of \texttt{CNOT} gates and single-qubit gates, see Fig. \ref{fig:figure}.

Here, $\ket{\psi}$ and $\ket{\phi}$ are arbitrary states, and the gates are given by 

\begin{align}
U_{1} &= 
\begin{pmatrix} 
1 & 0 \\
0 & e^{\frac{i\pi}{2n}}
\end{pmatrix} \nonumber
\\
R_{Y} \left(\frac{\pi}{4}\right) &= \frac{1}{\sqrt{2}}
\begin{pmatrix}
1 & -1 \\
1 & 1
\end{pmatrix} \nonumber
\\
R_{Y} \left(-\frac{\pi}{4}\right) &= \frac{1}{\sqrt{2}}
\begin{pmatrix}
1 & 1 \\
-1 & 1
\end{pmatrix} \nonumber
\\
R_{Z} (\eta) &=
\begin{pmatrix}
e^{-i\eta} & 0 \\
0 & e^{i\eta}
\end{pmatrix} \nonumber
\\
R_{Z} (-2\eta) &=
\begin{pmatrix}
e^{2i\eta} & 0 \\
0 & e^{2i\eta}
\end{pmatrix}
\end{align} 
where for $\alpha = \pi(\text{mod}\, 2\pi)$ the operator $U^{\alpha}$ can act as \texttt{$\mathtt{SWAP}$}$\ket{\phi}\ket{\psi} = \ket{\psi}\ket{\phi}, \forall \ket{\psi}\ket{\phi}$, while for $\alpha \sim 0$, $U^{\alpha}\equiv\mathbb{1}_{{SR}}$. Moreover, the closer $\eta$ is to $\pi/2$, the higher the probability is that $U^{\alpha}$ will act as the \texttt{$\mathtt{SWAP}$}, while for weak coupling the input state is only slightly modified towards the target reservoir state.

Using the two-body Heisenberg exchange operator $U^{\alpha}$ is advantageous for the practical implementation of the proposed protocol (\textit{e.g.} solid state quantum computing) as this engineered system-reservoir interaction driven stabilization of the input state onto a reservoir subspace is achieved by controlling only the coupling parameter $\alpha$ \cite{heisenberg}. This can be done via an external control field, \textit{e.g.} a global time-dependent magnetic field \cite{heisenberg2, heisenberg3}, eliminating the need for selective control over individual degrees of freedom \cite{overhead} and simplifying the architecture design.\footnote{This principle has been successfully applied for the engineering of effective ``always-ON" Hamiltonians in analogue quantum simulators \cite{qsimulators, qsimulators2}, leading to reduced computational cost and significantly improved system controllability.}\\
    
\section{Robustness of quantum homogenization}
In the context of approximate quantum error correction \cite{subsystem} and following the prescription of \cite{oreshkov}, we now demonstrate that such dissipative reservoir-based systems satisfy the Beny-Oreshkov generalization of the KL conditions for approximate quantum code correctability, herein viewed as a condition on the access to information by the environment. Namely, assuming there are some noise-induced non-unitary effects (as discussed below), we show that the associated information ``leakage`` is minimal and does not disturb the steady state of the quantum homogenizer.

Let us first remark that in the already introduced model, the system-reservoir interaction is unitary and purifies the homogenization process.\footnote{Where in practice, the presence of noise during the encoding of the input would induce some non-unitary propagation of the state to the environment $E$, such that the proposed collision model is the Stinespring dilation of the noisy process. In this case, such leakage of information into $E$ is described by the complementary channel $\widehat{\zeta}$, see Eq. (\ref{eq:complementary}).} Suppose we denote the complete operation as ${\mathcal{R}}\circ\zeta$, consisting of an encoding quantum channel $\zeta: S(\mathcal{H}_{{S}})\rightarrow S(\mathcal{H}_{{R}})$, representing the partial \texttt{$\mathtt{SWAP}$}, and a recovery channel $\mathcal{R}: S(\mathcal{H}_{{R}})\rightarrow S(\mathcal{H}_{{S}})$, representing its inverse, where $S(\mathcal{H}_{{S}})$ and $S(\mathcal{H}_{{R}})$ are the state space on the system and reservoir, respectively.\footnote{Ideally, the state is obtained by viewing $\zeta$ as the unitary partial \texttt{$\mathtt{SWAP}$} and discarding the environment $E$. As was shown in \cite{vlatko}, however, in practical setups complete tracing out of the degrees of freedom of $E$ is not possible, and some non-zero coupling between $R$ and $E$ remains.} Here, as is the case in any experimental setup, we consider the encoding channel $\zeta$ to be noisy in a sense that during the input-reservoir encoding process part of the information propagates into an environment $E\in \mathcal{H}_{E}$. Where for ease of notation we assume the noise elements $\mathcal{N}_{i}$ are absorbed in $\zeta$. Then, there exists a corresponding Stinespring dilation isometry \cite{dilation} $\mathcal{M}$ acting as $\mathcal{M}:\mathcal{H}_{{S}}\rightarrow\mathcal{H}_{{R}}\otimes\mathcal{H}_{E}$ which is unique up to local isomorphisms of $\mathcal{H}_{{E}}$. The isometry transforms the input state $\rho_{S}\in\mathcal{H}_{S}$ on $\mathcal{H}_{R}\otimes\mathcal{H}_{E}$ as $\mathcal{M}\rho_{S}\mathcal{M}^{\dagger}$.

We can now continue with the main task of characterizing the recoverability of the system information from the environment by employing the concept of \textit{complementary} channels which are known to have similar Stinespring representation \cite{oreshkov, comp}. This method is particularly useful since it relates distances between channels to distances between their \textit{complementary} versions. The encoding channel $\widehat{\zeta}:S(\mathcal{H}_{{S}})\rightarrow S(\mathcal{H}_{E})$, complementary to $\zeta$, is defined as

\begin{equation}
\label{eq:complementary}
\widehat{\zeta}(\rho_{{S}}) = \text{tr}_{R}\left(\mathcal{M}\rho\mathcal{M}^{\dagger}\right)
\end{equation}
where $\widehat{\zeta}$ specifies how much information from the input system $\mathcal{H}_{{S}}$ has leaked into the environment. Because of that, all complementary channels are also unique in the same restricted sense as the isometry $\mathcal{M}$, and are associated with choices of states in the environment.\footnote{Likewise, complementary versions can be defined for all quantum channels. For instance, $\widehat{\mathcal{F}}=\text{tr}$ is the complementary to $\mathcal{F}=\mathbb{1}$.} In this setup, the generalized KL conditions allow us to test whether an approximate recovery channel exists, as measured by the fidelity-based Bures distance $\mathcal{B}$, studied in \cite{oreshkov, measures}, where we consider the following definition as a necessary and sufficient condition\\ 

\textbf{Definition \hypertarget{definition}{1}.}
An encoding map $\mathcal{M}$ defines a code which is $\delta$-correctable under the channel $\zeta$, where a recovery channel $\mathcal{R}$ as in $({\mathcal{R}}\circ\zeta,\mathcal{F}) = 1-\delta$ exists iff a \textit{complementary} channel $\widehat{\mathcal{R}}$ as in $(\widehat{\zeta},\widehat{\mathcal{R}\circ{\mathcal{F}}}) = 1-\delta$ exists, where $\mathcal{F}$ is a target channel (\textit{e.g.} $\mathcal{F} = \mathbb{1}$) and $\delta \in [0,1]$.\\~\\
We can now restate the above definition in terms of the diamond norm which more accurately captures the presence of entanglement with an ancillary system \cite{diamond} as

\begin{equation}
\label{eq:bures}
\|{\mathcal{R}}\,\circ\,\zeta - \mathcal{F}\|_{\diamond}\leq\delta
\end{equation}
for $\delta\ll 1$. Note the generalized KL conditions are broadly applicable for subsystem and algebraic codes and the recovery operation may be taken with respect to an arbitrary target channel \cite{22}. To simplify things, we can convert Definition \hyperlink{definition}{1} into a condition on the existence of the complementary channel $\widehat{\zeta}$: If $\forall\rho_{{S}}\in S(\mathcal{H}_{{S}})$ there exists a channel with approximately constant output to the environment, such as $\widehat{\zeta}(\rho_{{S}})\approx\varphi\text{tr}(\rho_{{S}})$ (where $\varphi$ is an arbitrary output state), then ${\mathcal{R}}$ must exist such that $\text{max}_{{\mathcal{R}}}({\mathcal{R}}\circ\zeta,\xi)\leq\delta$ for $\delta\ll 1$, where in terms of (\ref{eq:bures}) $\delta\leq\text{min}_{{\mathcal{R}}}\mathcal{B}({\mathcal{R}}\circ\zeta,\mathcal{F})\leq\text{min}_{\widehat{\mathcal{R}}}\mathcal{B}(\widehat{\zeta},\widehat{\mathcal{R}\circ\mathcal{F}})$.

Recall that in our setup, the encoding channel $\zeta$ (specified by the isometry $\mathcal{M}$) is noisy with elements $\mathcal{N}_{i}$. In this sense, as it was pointed out in \cite{oreshkov}, the condition on the existence of an approximately constant output channel $\widehat{\zeta}$ depends on whether there exists a set of coefficients $\lambda_{ij}$ (\textit{i.e.} density operator components), such that $\mathcal{M}^{\dagger}\mathcal{N}_{i}^{\dagger}\mathcal{N}_{j}\mathcal{M}=\lambda_{ij}\mathbb{1}$ is satisfied, where $\lambda_{ij}=\bra{i}\varphi\ket{j}$ for some output state $\varphi$. Assuming $\widehat{\mathcal{F}}$ is a projection ($\widehat{\mathcal{F}} = \widehat{\mathcal{F}}^{2}$), a large family of such coefficients is known to exist, such as $\lambda_{ij} = \text{tr}(\beta\mathcal{N}_{i}^{\dagger}\mathcal{N}_{j})$ for some state $\beta$. Moreover, given the channel $\zeta$, its complementary $\widehat{\zeta}$, and the corresponding isometry $\mathcal{M}$ are defined as above, let $\mathcal{P}$ be a channel which maps all logical states to some constant state in $S(\mathcal{H}_{E})$. Concretely, $\mathcal{P}$ maps the input states as

\begin{equation}
\label{eq:leaking}
\mathcal{P}:S(\mathcal{H}_{S})\rightarrow {\sigma}
\end{equation}
where $\sigma\in S(\mathcal{H}_{E})$ denotes a constant state in the environment $E$, corresponding to a subsystem $\tilde{E}\subset E$ of dimension at most $k$ (denoted below as $\tilde{E}^{\sigma}_{k}$), as it is naturally expected by the quantum homogenization dynamics.\footnote{Note that (i) $\mathcal{P}$ is absorbed in $\widehat{\zeta}$ similar to how $\mathcal{N}$ is absorbed in $\zeta$, where for ease of notation $\mathcal{P}$ is dropped throughout the paper, and (ii) $\rho_{{S}}$ is assumed to be simple one-dimensional logical state. Otherwise, $||\rho_{{S}}|| \geq k$ can quickly saturate the channel capacity.} From the information-disturbance theorem \cite{info} we know that

\begin{equation}
\label{eq:forgetful}
\parallel\widehat{\zeta}-\mathcal{P}\parallel  \leq \parallel\mathcal{R}\circ\zeta-\xi^{(0)}\parallel    \
\end{equation}
where $\xi^{(0)}$ is the initial reservoir state as defined earlier. Notice that here, the theorem states that if a channel $\mathcal{R}$ exists, such that it can recover $\zeta$ arbitrarily well, the r.h.s norm of (\ref{eq:forgetful}) will be small. Meaning that, short of a macroscopic measurement-induced disturbance, the system leaks almost no information to the environment, as $\widehat{\zeta}$ is well approximated by $\mathcal{P}$.

Now, suppose $\zeta$ is such that its complementary $\widehat{\zeta}$ is approximately $k$-forgetful with an uncertainty $\delta$ \cite{hayden}, meaning that the channel output is to a subspace of dimension less than or equal to $k$

\begin{equation}
\label{eq:forgetful2}
\parallel\widehat{\zeta}-\mathcal{P}\parallel^{(k)}_{\diamond} \leq \delta
\end{equation}
where $\delta \ll 1$ and $k$ denotes the maximal dimension of the encoded subspace in $E$. Evidently, the complementary channel $\widehat{\zeta}: \rho_{{S}}\rightarrow \tilde{E}^{\sigma}_{k}$ is simply the restriction of $\zeta$ to act on $S(\mathcal{H}_{\tilde{E}})$ for any arbitrary subsystem $\tilde{E}\subset E$ of dimension at most $k$. From (\ref{eq:leaking}--\ref{eq:forgetful2}), and given that $\zeta$ is a channel with fixed single-point output, its complementary $\widehat{\zeta}$ is also an approximately constant output channel $\widehat{\zeta}(\rho_{{S}})\approx\varphi\text{tr}(\rho_{{S}})$ for $\forall\rho_{{S}}\in S(\mathcal{H}_{{S}})$ and some fixed state $\varphi \in \mathcal{H}_E$. Thus, it straightforwardly follows from the subsystem decoupling theorem \cite{hayden} (which formalizes the relation between approximate correctability and the forgetfulness of $\widehat{\zeta}$ to an environment subspace) that for all subsystems $\tilde{E}\subset E$ of dimension $\leq k$, where $k\ll|R|$, an approximate recovery channel $\mathcal{R}$ must exist (with precision of $2\sqrt{2\delta}$ as a consequence of \cite{info}), such that it satisfies Definition \hyperlink{definition}{1} \cite{oreshkov}. Notice that an essential feature of the protocol is the convergence dynamics towards a desired reservoir target state for all logical states. Here, for all logical states the dynamics asymptotically outputs a desired fixed state, such that $\forall\rho_{{S}}\in S(\mathcal{H}_{{S}})$ the complementary channels $\widehat{\mathcal{R}\circ\mathcal{F}}(\rho_{{S}}) = \xi^{(0)}$ (in the neighborhood of $\delta$) exist.

Therefore, although in practical setups perfect tracing over the environment degrees of freedom is not possible as some non-zero coupling remains, we have demonstrated that the "leaked" information is minimal and recoverable, and does not disturb the system. Hence, the desired quantum steady state remains stabilized in the reservoir subspace as $\widehat{\zeta}$ is well approximated by $\mathcal{P}$. Of course, implementing such a recovery protocol would, in addition to hardware dependence, also necessitate having a detailed information about the contributions to the noise due to non-zero correlations with the environment. 

\section{Conclusions}
We have presented a new formulation of the quantum homogenization protocol with two-body Heisenberg-type exchange interactions which satisfies the generalized KL conditions for approximate error correctability. The proposed protocol contributes to the recently established research on dissipation-driven information protection \cite{dissipative, dissipative2, dissipative3} as it presents a physically-implementable platform whose naturally convergent dynamics asymptotically drives it to a desired steady state manifold without the need for manipulating individual degrees of freedom. Instead, it only requires control over an external field to tune the exchange interactions, thus eliminating the hardware overhead that many current methods suffer from \cite{overhead}.

The realization of such systems, capable of robust state preparation and manipulation of arbitrary states, is essential for the success of quantum hardware and is of broad interest to several communities. The ability to deterministically generate entangled states between multidimensional quantum systems (\textit{e.g.} qudits) has many practical applications due to their demonstrated robustness compared to qubits, such as quantum key distribution \cite{qkey} and hybrid quantum information processing \cite{hybrid}. Moreover, such reservoir-based systems can be further extended to more complex schemes such as topological error correction \cite{sensing}, as well as be utilized for the development of quantum simulators \cite{simulator}. Interestingly, the possible applications of the current framework as a quantum cloning protocol or a quantum safe were explored in \cite{6, qcloning}, while its utility in spin chain quantum communication protocols in \cite{qcomm}. From a theoretical perspective, on the other hand, such systems are broadly studied in resource theory, where they are referred to as \textit{catalysts} \cite{resource}.

Ultimately, this line of work can not only help in refining existing quantum hardware and quantum computation protocols, but also lead to new insights into quantum information theory and quantum thermodynamics.

\section*{Acknowledgments}

V.V. thanks the Oxford Martin School, the John Templeton Foundation, the Gordon and Betty Moore Foundation, and the EPSRC (UK). The authors thank Maria Violaris for the valuable discussions.

\end{document}